\RequirePackage{lineno}
\documentclass[final,5p,times,twocolumn,compress]{elsarticle}

\usepackage{orcidlink}

\newcommand{\BR}{{\cal B}}

\usepackage{amsmath}

\usepackage{graphicx}
\usepackage{epsfig}
\usepackage{dcolumn}
\usepackage{bm}
\usepackage{amsmath}
\usepackage{color}
\usepackage{color,xcolor}
\usepackage{lineno}
\usepackage{threeparttable}
\usepackage{rotating}

\begin{document}

\begin{frontmatter}

\title{\quad\\[0.0cm] Observation of \boldmath{$\Omega(2012)^- \to \Xi(1530)\bar{K}$} and measurement of the effective couplings of \boldmath{$\Omega(2012)^-$ to $\Xi(1530)\bar{K}$} and \boldmath{$\Xi\bar{K}$}
}

\vspace{-10cm}
\author{(The Belle Collaboration)\\S.~Jia\,\orcidlink{0000-0001-8176-8545}}
\author{C.~P.~Shen\,\orcidlink{0000-0002-9012-4618}}
\author{C.~Z.~Yuan\,\orcidlink{0000-0002-1652-6686}}

  \author{J.~K.~Ahn\,\orcidlink{0000-0002-5795-2243}}
  \author{H.~Aihara\,\orcidlink{0000-0002-1907-5964}}
  \author{D.~M.~Asner\,\orcidlink{0000-0002-1586-5790}}
  \author{H.~Atmacan\,\orcidlink{0000-0003-2435-501X}}
  \author{R.~Ayad\,\orcidlink{0000-0003-3466-9290}}
  \author{S.~Bahinipati\,\orcidlink{0000-0002-3744-5332}}
  \author{Sw.~Banerjee\,\orcidlink{0000-0001-8852-2409}}
  \author{J.~Bennett\,\orcidlink{0000-0002-5440-2668}}
% ...

\begin{abstract}
\linenumbers
Using $\Upsilon(1S)$, $\Upsilon(2S)$, and $\Upsilon(3S)$ data collected by the Belle detector, we discover a new three-body decay, $\Omega(2012)^-\to\Xi(1530)\bar K\to\Xi\pi\bar K$, with a significance of 5.2~$\sigma$.
The mass of the $\Omega(2012)^-$ is $(2012.5\pm0.7\pm0.5)$ MeV and its effective couplings to $\Xi(1530)\bar{K}$ and $\Xi\bar{K}$ are $(39^{+31}_{-39}\pm9)\times10^{-2}$ and $(1.7\pm0.3\pm0.3)\times10^{-2}$, where the first uncertainties are statistical and the second are systematic.
The ratio of the branching fraction for the three-body decay to that for the two-body decay to $\Xi\bar{K}$ is $0.99\pm0.26\pm0.06$, assuming isospin symmetry.
\end{abstract}


\end{frontmatter}

\section{Introduction}

In the last twenty years, physicists have discovered new hadrons that
are difficult to explain with the conventional quark model~\cite{639,015003,873}.
Some of these, like the $X(3872)$, $Z_c(3900)$, and $T_{cc}(3875)$ mesons and
the $P_c(4312)$ baryon, are close to the mass thresholds of
pairs of conventional hadrons. Many authors propose that they are
hadronic molecules~\cite{873,015004}.~All these hadrons have two heavy quarks, but the study of such hadrons containing only light quarks is rather
limited. The $\Lambda(1405)^0$, close to the $pK^-$ threshold, is widely regarded
as an $N\bar K$ molecule~\cite{132002,1593,55,38}. New measurements of the properties of
similar hadrons will let us refine the model of hadronic molecules and
improve our understanding of the strong interaction between quarks and
gluons~\cite{1095,076301}.

One such hadron is the $\Omega^-$, which with three strange quarks is the strangest baryon.
Its excited states have been difficult to find.
The \emph{Review of Particle Physics}~\cite{PDG} lists only four excited $\Omega^-$ baryons: $\Omega(2012)^-$, $\Omega(2250)^-$, $\Omega(2380)^-$, and $\Omega(2470)^-$. The last three were established four decades ago~\cite{33,579,799}.
The $\Omega(2012)^-$ was first observed in 2018 by the Belle experiment in decays to $\Xi^0K^-$ and $\Xi^-{\bar K}^0$
in $e^+e^-$-collision data near the $\Upsilon(1S)$, $\Upsilon(2S)$, and $\Upsilon(3S)$ resonances, who reported it to have a mass of $(2012.4\pm0.9)$ MeV~\cite{unit} and a width of ($6.4^{+3.0}_{-2.7})$ MeV~\cite{052003}.

Theorists are strongly interested in understanding the $\Omega(2012)^-$~\cite{034004,894,114023,016002,10427,025202,063104,014025,014031,04458,55v2,054009,056013,857,076012,025001,13396,094016,361,074025}.
There are mainly two interpretations: a standard baryon~\cite{034004,894,114023,016002,10427,025202,063104,014025,014031,04458} and a $\Xi(1530)\bar{K}$ molecule~\cite{55v2,054009,056013,857,076012,025001,13396,094016,361,074025}.
A large rate for $\Omega(2012)^-$ to $\Xi(1530)\bar K$ was predicted in the molecule scenario~\cite{55v2,054009,056013,857,076012}---18\% to 86\% of that for decay to $\Xi\bar K$~\cite{054009,056013}. Therefore, measuring the branching fraction for $\Omega(2012)^-\to \Xi(1530)\bar{K}\to \Xi\pi\bar{K}$ may inform us about the $\Omega(2012)^-$'s internal structure.

We report a search for $\Omega(2012)^-\to\Xi(1530)\bar K$ using $\Upsilon(1S)$, $\Upsilon(2S)$, and $\Upsilon(3S)$ data collected by the Belle experiment~\cite{Belle1,Belle2} at the KEKB energy-asymmetric $e^+e^-$ collider~\cite{KEKB1,KEKB2}. We measure its branching fraction with respect to that for $\Omega(2012)^-\to\Xi\bar K$. Charge-conjugate modes are included throughout.

\section{Comparison with previous analysis}

Belle previously searched for $\Omega(2012)^-\to\Xi(1530)\bar{K}$, but observed no signal and set an upper limit at 90\% confidence on the branching-fraction ratio~\cite{032006},

\vspace{-0.2cm}
\begin{equation}\label{ratio}
{\cal R}^{\Xi\pi{\bar K}}_{\Xi {\bar K}}\equiv\frac{\BR(\Omega(2012)^-\to \Xi(1530)\bar{K})}{\BR(\Omega(2012)^-\to \Xi \bar{K})}<11.9\%.
\end{equation}

That analysis was based on an inaccurate model of the decay, which led to selection criteria that accepted more background events than expected, a misestimation of the detection efficiency, and an uncertain signal yield. Our new analysis is based on a more accurate model of the decay and uses different selection criteria, accounting for the fact that the $\Xi(1530)$ is produced below threshold. We use this model, including a more detailed resonance shape and a previously neglected three-body phase-space factor, to fit for the signal yield.
The distribution of $M(\Xi^-\pi^+)$ in the previous model peaks more narrowly and at higher masses than the distribution in the updated model (Fig.~\ref{MXipi}). This leads us to determine different selection criteria than the previous analysis.
We further improve upon the previous analysis by vetoing events in which the $K^-$ may come from a $\phi$ instead of an $\Omega^-$.

These improvements increase the signal yield and greatly change the results for ${\cal R}^{\Xi\pi{\bar K}}_{\Xi {\bar K}}$, which supersede those of \cite{032006}. 
In the new decay model, the mass of the $\Omega(2012)^-$ and its couplings to $\Xi\pi\bar K$ and $\Xi \bar K$ are free parameters.
We measure these parameters, superseding the results reported in~\cite{052003}.

\section{Data sample and Belle detector}

The data used in this analysis were taken at the $\Upsilon(1S)$, $\Upsilon(2S)$, and $\Upsilon(3S)$ resonances with integrated luminosities of 5.7, 24.9, and 2.9 fb$^{-1}$, corresponding to 102 million $\Upsilon(1S)$, 158 million $\Upsilon(2S)$, and 12 million $\Upsilon(3S)$ events.
Since the $\Upsilon(1S)$, $\Upsilon(2S)$, and $\Upsilon(3S)$ states decay via three gluons or two gluons and a
photon, they produce more baryons than continuum $e^+e^-\to q\bar q$ ($q$ = $u$, $d$, $s$, $c$)~\cite{2161,177,012005,64301}, so the signal-to-background ratio in on-resonance $\Upsilon(1S)$, $\Upsilon(2S)$, and $\Upsilon(3S)$ data is larger than those in continuum and heavier $\Upsilon$ data.

The Belle detector is a large solid-angle magnetic spectrometer consisting of a silicon vertex detector, a 50-layer central drift chamber (CDC), an array of aerogel threshold Cherenkov counters (ACC), a barrel-like arrangement of time-of-flight scintillation counters (TOF), and an electromagnetic calorimeter (ECL) comprised of CsI(Tl) crystals located inside a superconducting solenoid coil that provided a 1.5 T magnetic field and an iron flux return placed outside the coil instrumented to detect $K^0_L$ and identify muons.

We use simulated events to determine detection efficiencies, signal shapes, and selection criteria. We use {\sc evtgen}~\cite{EVTGEN} to generate simulated events. 
We use a Flatt$\acute {\rm e}$-like function~\cite{094028} to simulate $\Omega(2012)^-\to \Xi(1530)\bar{K}\to \Xi \pi \bar K$.
We determine preliminary selection criteria assuming the $\Omega(2012)^-$ mass is $2012.4$ MeV~\cite{052003} and its couplings to $\Xi \bar K$ and $\Xi \pi\bar K$ are 0.01 and 0.1. We then update the model to the values we find in data and re-optimize our selection criteria. We repeat this several times until the values stabilize.
Simulated samples of inclusive $\Upsilon(1S)$, $\Upsilon(2S)$, and $\Upsilon(3S)$ decays and $e^+e^-\to q\bar q$ ($q=u,d,s,c$) with four times the luminosity as the real data are produced to check for possible peaking backgrounds~\cite{107540}. {\sc geant3}~\cite{1984} is used to simulate detector response.

\section{Selection criteria}

We search for the $\Omega(2012)^-$ in the $\Xi^0K^-$, $\Xi^-\bar K^0$, $\Xi^-\pi^+K^-$, $\Xi^-\pi^0\bar K^0$, $\Xi^0\pi^-\bar K^0$, and $\Xi^0\pi^0K^-$ final states, where the three-body final states are used to look for the $\Xi(1530)$ resonance via its decay to $\Xi\pi$. We use the same selection methods and criteria for $\pi^+$, $K^-$, $p$, $\pi^0$, $\bar K^0$, $\Lambda^0$, $\Xi^-$, and $\Xi^0$ as in Refs.~\cite{052003,032006}.
We require the mass of $\Xi\bar K$ and $\Xi\pi\bar K$ be less than 2200 MeV.

We veto $K^-$ that may come from decay of a $\phi$ instead of an $\Omega(2012)^-$ by rejecting events fulfilling $|M(K^-K^+)-m_{\phi}|$ $<$ 10 MeV for any $K^+$ in the rest of the event, where the $K^+$ must be identified as a kaon with the same requirements as for the $K^-$.
This window covers five units of the mass resolution and rejects 96\% of $\phi$-induced backgrounds.

\section{$\Xi\pi$ distributions}

\begin{figure}[t]
\centering
\includegraphics[width=7cm]{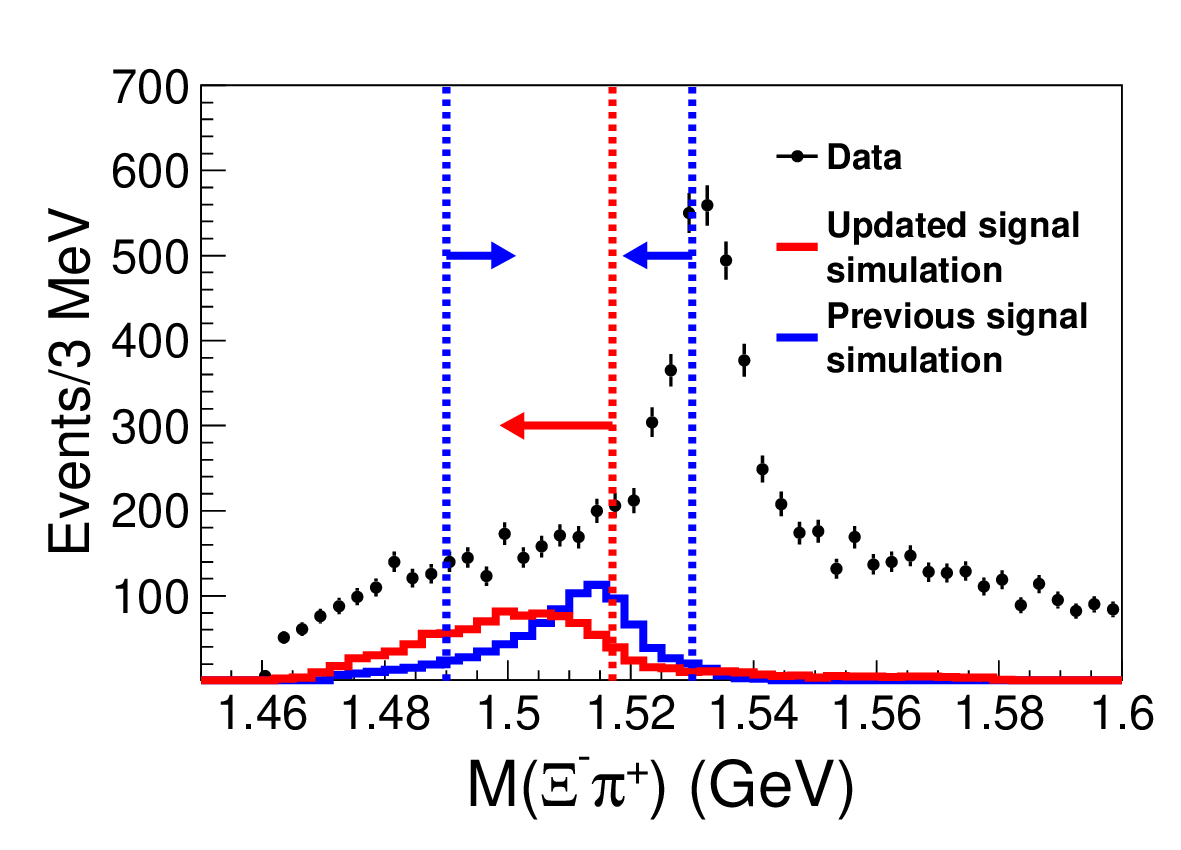}
\caption{The $\Xi^-\pi^+$ mass distributions in $\Xi^-\pi^+K^-$ events in data (points) and in simulations using the models of the previous analysis~\cite{032006} (solid blue line) and this work (solid red line), with each simulation normalized to have an exaggerated ${\cal R}^{\Xi\pi{\bar K}}_{\Xi {\bar K}}$ value of 4.
The red arrow shows the mass requirement of this analysis; the blue arrows show the mass requirement of the previous analysis~\cite{032006}.}\label{MXipi}
\end{figure}

Fig.~\ref{MXipi} shows the $\Xi^-\pi^+$ mass distributions in $\Xi^-\pi^+K^-$ events in data and in simulation using the model of the previous analysis~\cite{032006} and our updated model. Each simulated distribution is normalized to have an exaggerated ${\cal R}^{\Xi\pi{\bar K}}_{\Xi {\bar K}}$ value of 4 so that it is visible in comparison to the data.
In this updated work, we require $M(\Xi^-\pi^+)$ $<$ 1517 MeV, shown by the red arrow, which maximizes $S/\sqrt{S+B}$, where $S$ and $B$ are the numbers of signal and background events according to simulation, assuming ${\cal R}^{\Xi\pi\bar K}_{\Xi\bar K} = 1$ and isospin symmetry relates the branching fractions for the different $\Xi\pi\bar K$ final states to each other.
This retains 79\% of signal candidates and removes large backgrounds from $\Xi(1530)$ not produced from $\Omega(2012)^-$ and from $\Xi^-\pi^+$ not produced from $\Xi(1530)$. These backgrounds cause a large peak at the $\Xi(1530)^0$ mass in the $\Xi^-\pi^+$ mass distribution. 
Signal events peak below the $\Xi(1530)$ mass because the $\Xi(1530)$ is produced below threshold in $\Omega(2012)^- \to \Xi(1530) \bar K$.
We apply the same $M(\Xi\pi)$ requirement for all three-body channels, optimizing it from the $\Xi^-\pi^+K^-$ channel since it has the largest signal yield, according to isospin symmetry, and largest signal-to-background ratio, owing to it having the simplest final state to detect.

\begin{figure*}[t]
\centering
\includegraphics[width=5.5cm]{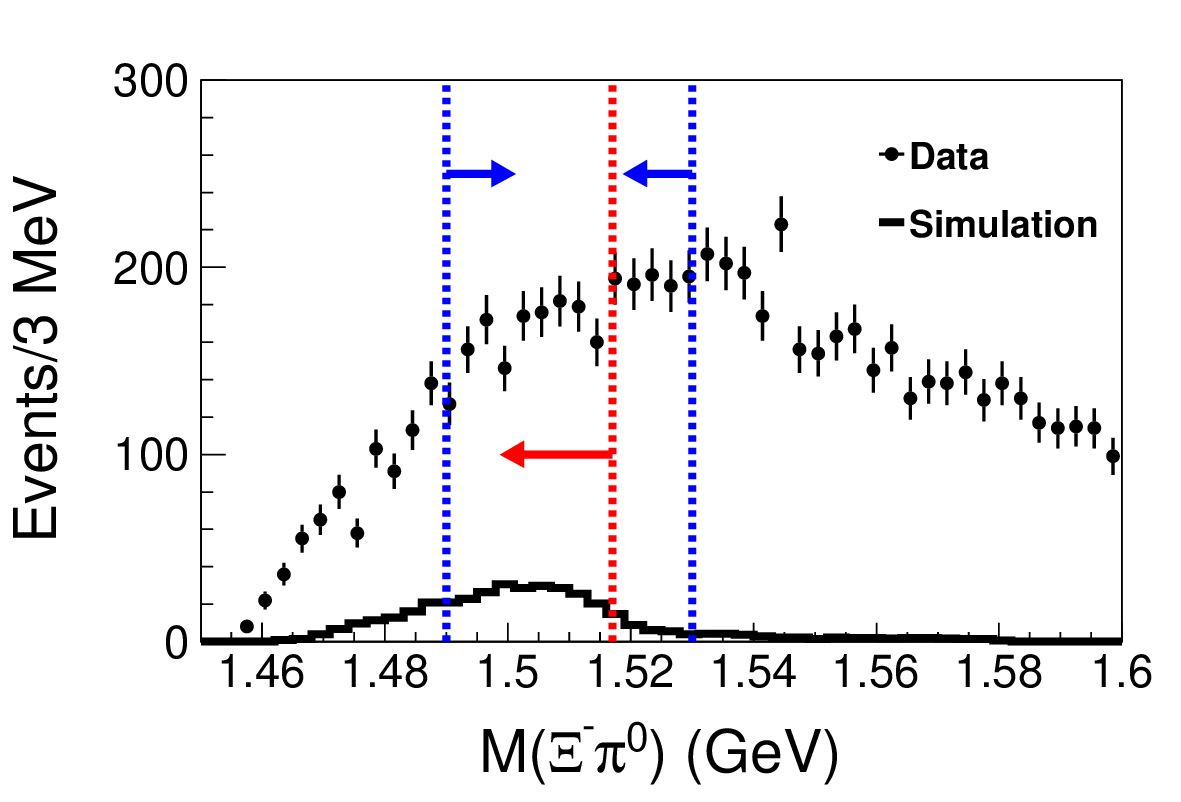}
\put(-122,85){\bf (a)}
\includegraphics[width=5.5cm]{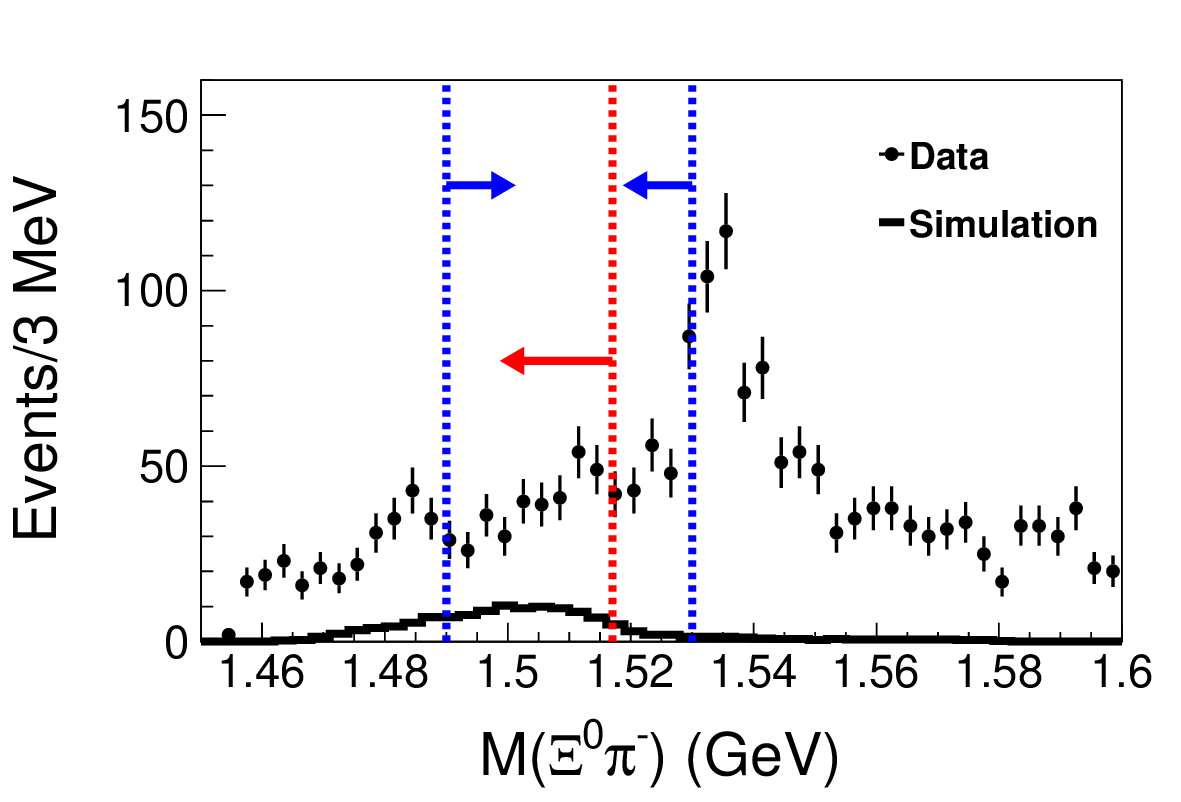}
\put(-122,85){\bf (b)}
\includegraphics[width=5.5cm]{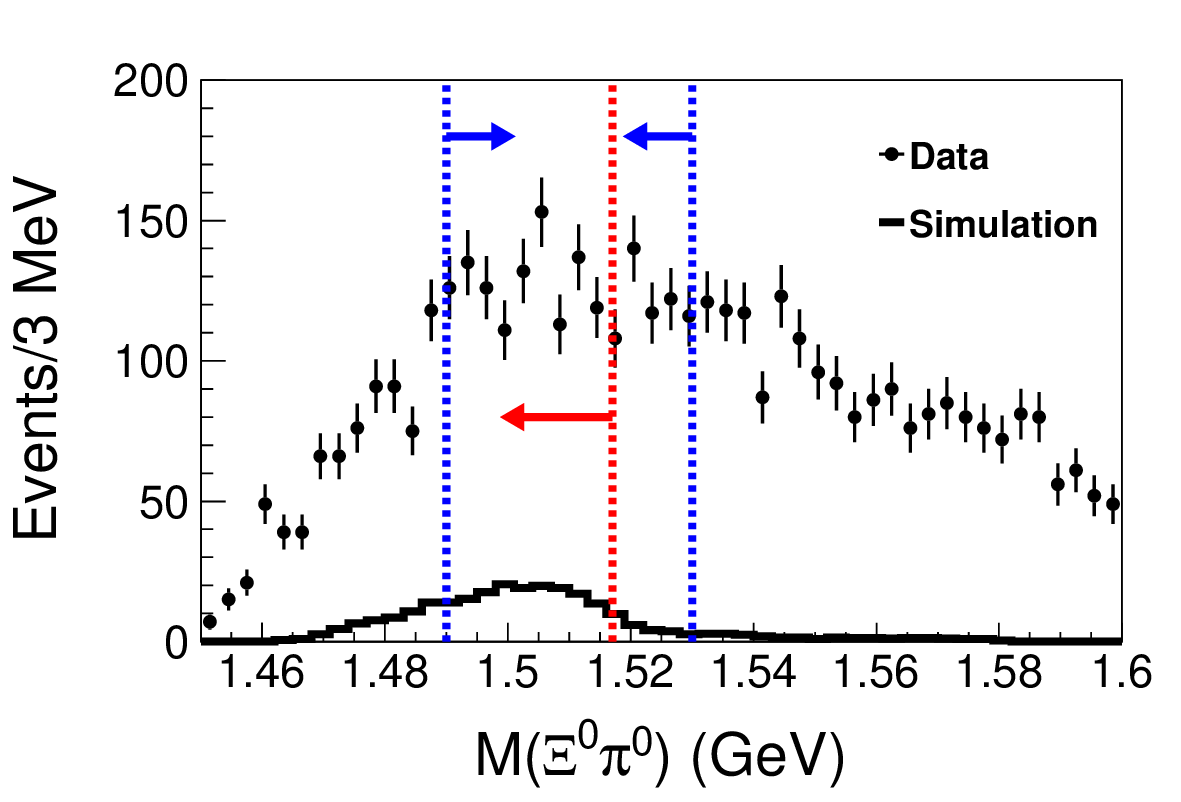}
\put(-122,85){\bf (c)}
\caption{
The (a) $\Xi^-\pi^0$, (b) $\Xi^0\pi^-$, and (c) $\Xi^0\pi^0$ mass distributions in $\Xi\pi\bar K$ events in the data and simulated $\Omega(2012)^-\to \Xi(1530)^-K^0_S\to \Xi^-\pi^0K^0_S$, $\Omega(2012)^-\to \Xi(1530)^-K^0_S\to \Xi^0\pi^-K^0_S$, and $\Omega(2012)^-\to \Xi(1530)^0K^-\to \Xi^0\pi^0K^-$ data.
The numbers of simulated signal events are scaled arbitrarily. The red arrows show the mass requirements of this analysis; the blue arrows show the mass requirements of the previous analysis~\cite{032006}.}\label{MXipi2}
\end{figure*}

The previous analysis required $M(\Xi\pi)$ $\in$ $[1490,1530]$\ MeV \cite{032006}, shown by the blue arrows in Fig.~\ref{MXipi}. This removed the 23\% of signal candidates that are below 1.49~GeV and accepted a large number of background events close to the $\Xi(1530)$ mass.
The updated $\Xi^-\pi^+$ mass requirement improves $S/\sqrt{S+B}$ by a factor of 1.4.
Fig.~\ref{MXipi2} shows the invariant mass distributions for $\Xi\pi$ for the final states $\Xi^-\pi^0\bar K^0$, $\Xi^0\pi^-\bar K^0$, and $\Xi^0\pi^0K^-$ and the previous and new mass requirements.

\section{$\Xi\pi\bar K$ and $\Xi\bar K$ distributions}

\begin{figure*}[t]
\centering
\includegraphics[width=5.5cm]{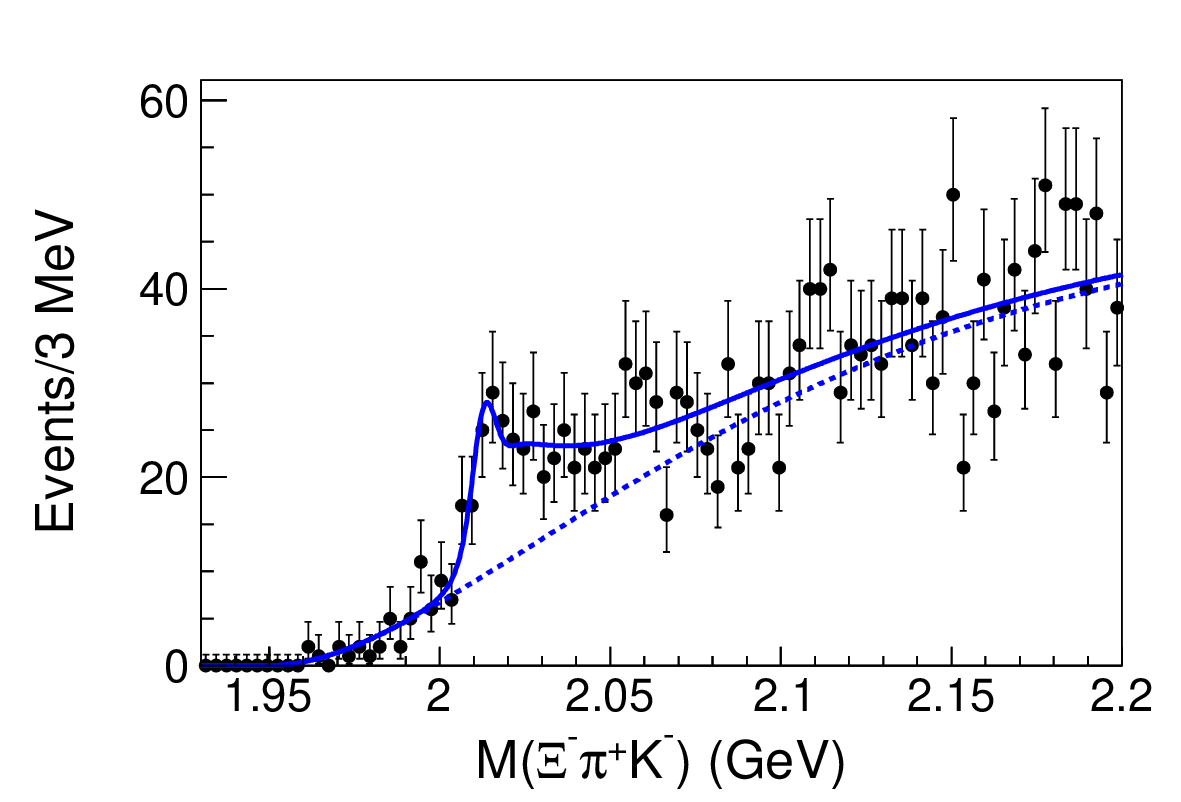}
\put(-122,85){\bf (a)}
\hspace{0.1cm}
\includegraphics[width=5.5cm]{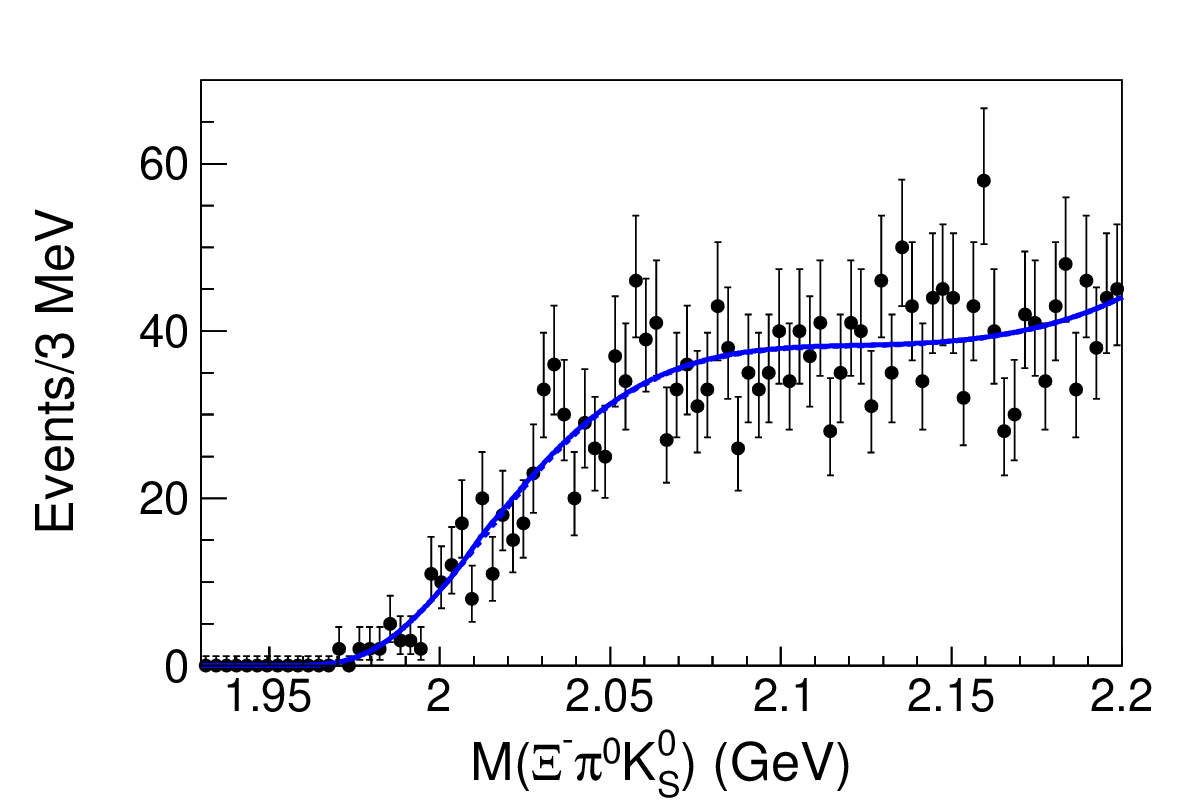}
\put(-122,85){\bf (b)}
\hspace{0.1cm}
\includegraphics[width=5.5cm]{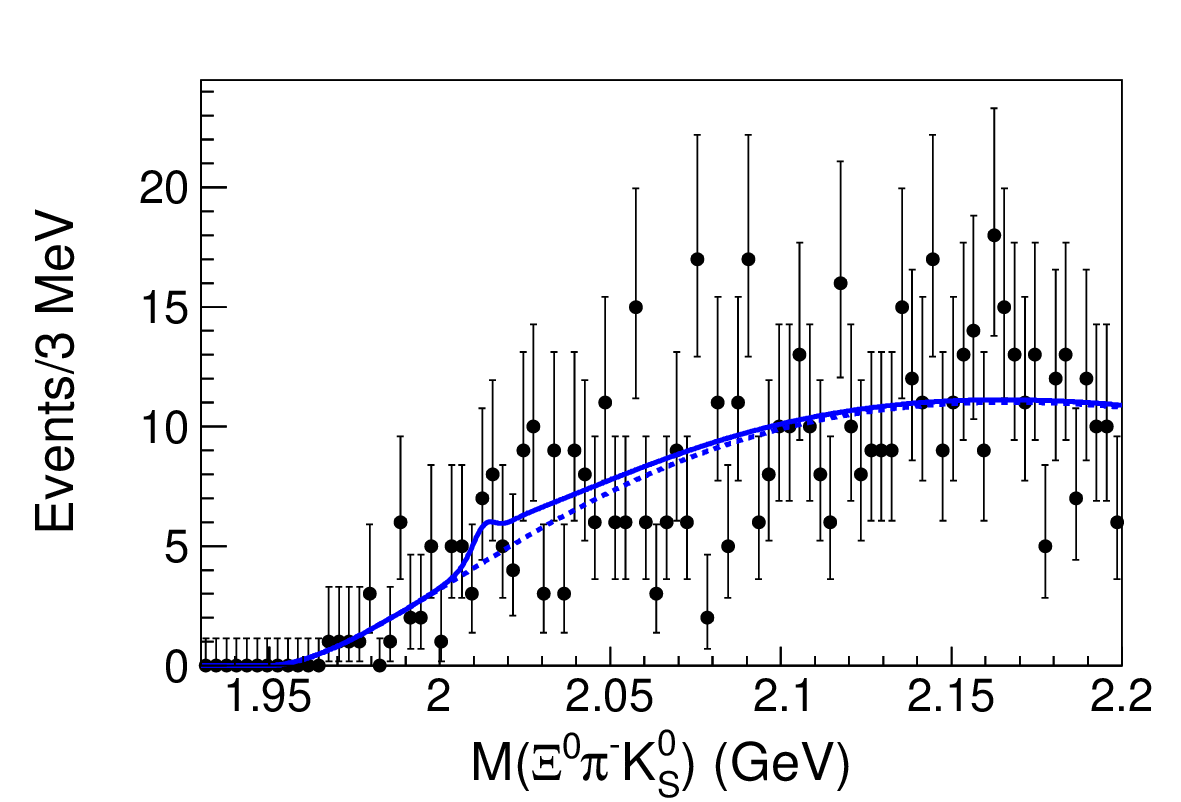}
\put(-122,85){\bf (c)}

\includegraphics[width=5.5cm]{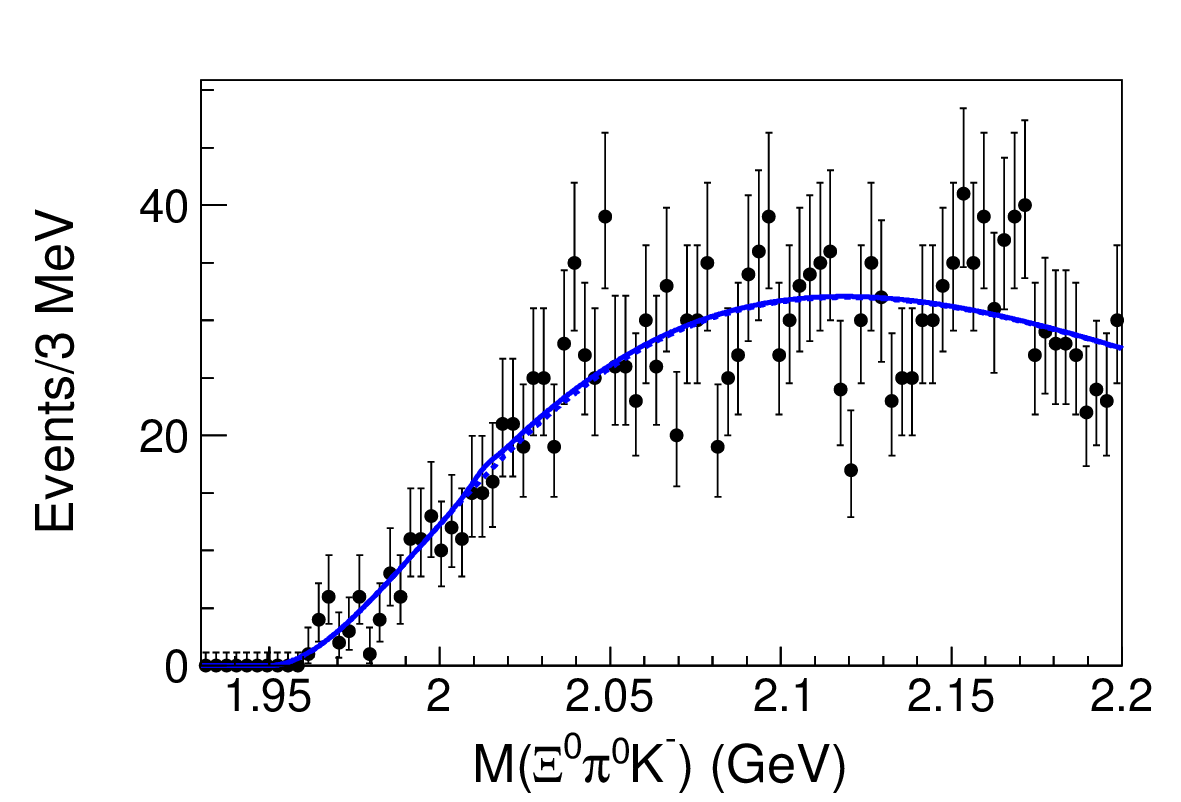}
\put(-122,85){\bf (d)}
\hspace{0.1cm}
\includegraphics[width=5.5cm]{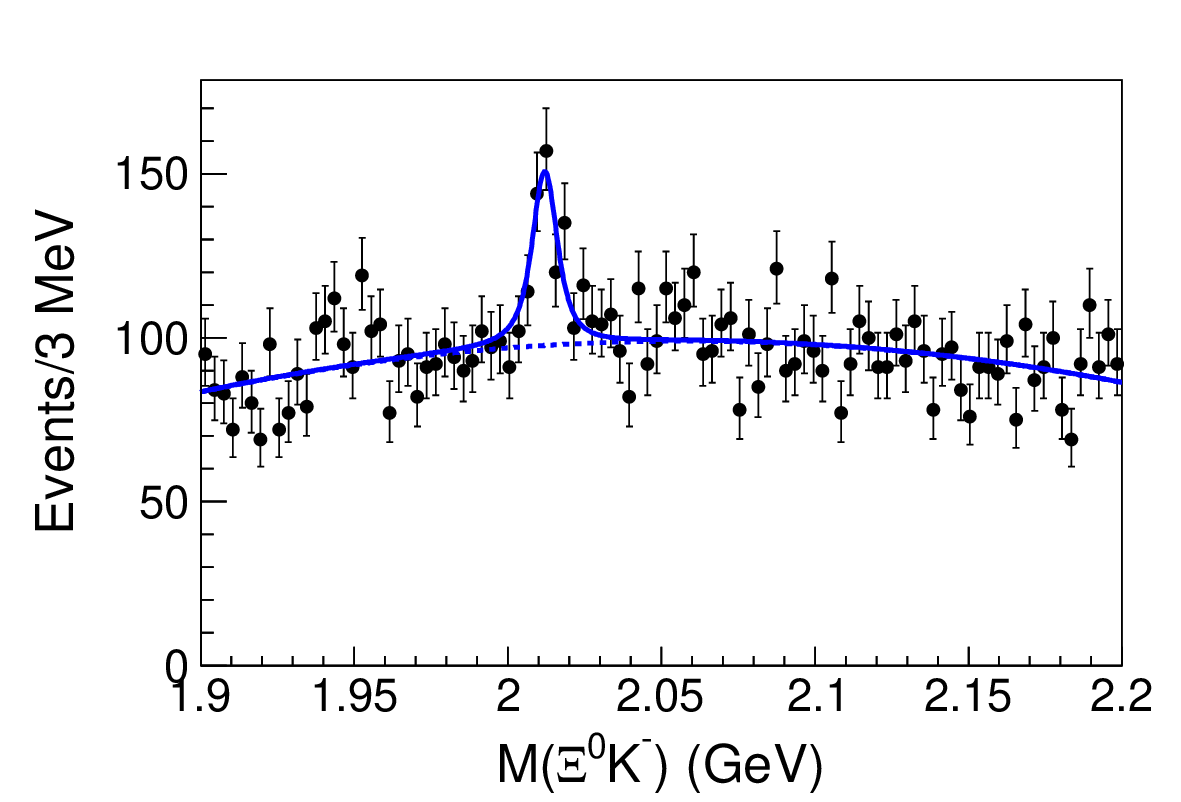}
\put(-122,85){\bf (e)}
\hspace{0.1cm}
\includegraphics[width=5.5cm]{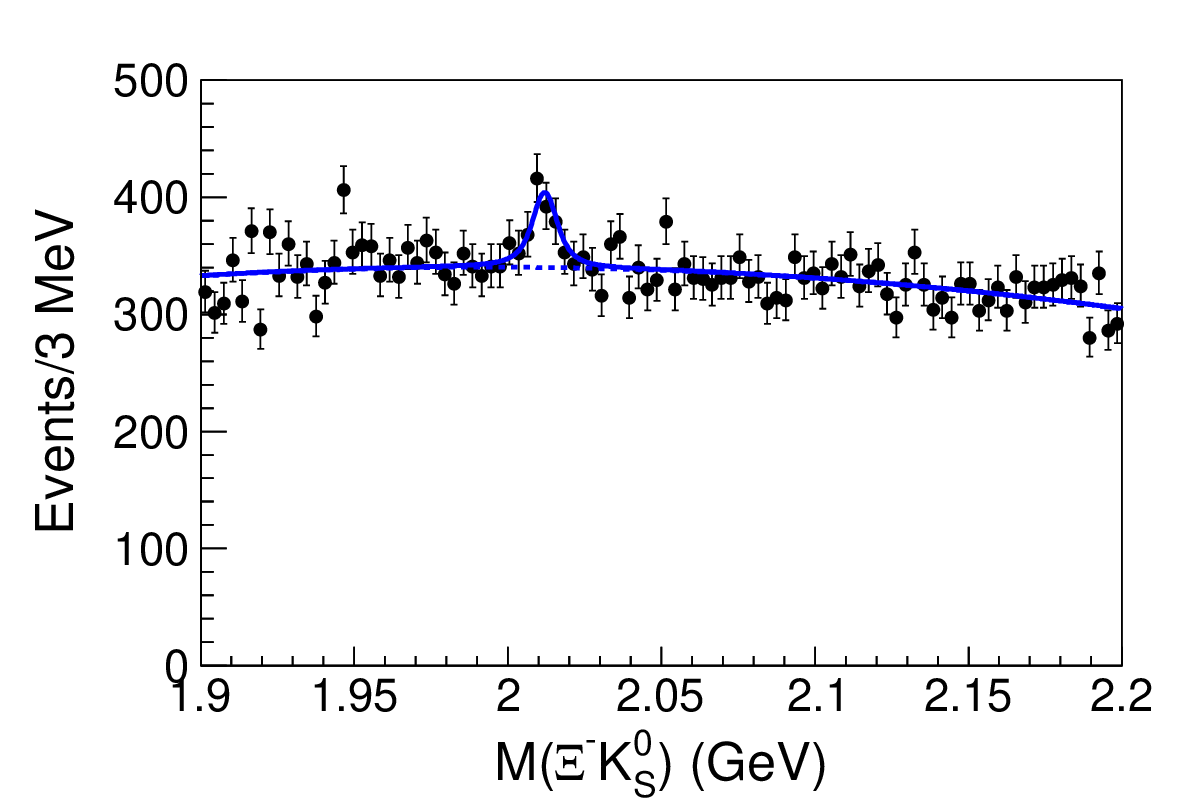}
\put(-124,85){\bf (f)}
\caption{The (a) $\Xi^-\pi^+K^-$, (b) $\Xi^-\pi^0K^0_S$, (c) $\Xi^0\pi^-K^0_S$, (d) $\Xi^0\pi^0K^-$, (e) $\Xi^0K^-$, and (f) $\Xi^-K^0_S$ mass distributions in data. The solid curves are the best fits and the dashed curves are backgrounds.}\label{fig2}
\end{figure*}

Fig.~\ref{fig2} shows the $\Xi\pi \bar K$ and $\Xi\bar K$ mass distributions in data after the all criteria.
No peaking backgrounds are found in simulated events.
To determine the $\Omega(2012)^-$ branching fractions and resonance parameters,
we fit simultaneously to the binned $\Xi^-\pi^+K^-$, $\Xi^-\pi^0K^0_S$, $\Xi^0\pi^-K^0_S$, $\Xi^0\pi^0K^-$, $\Xi^0K^-$, and $\Xi^-K^0_S$ mass distributions.

Under isospin symmetry, the branching fractions for $\Omega(2012)^-$ decay to $\Xi^-\pi^+K^-$ and $\Xi^0\pi^-\bar K^0$ are equal, the branching fractions for $\Omega(2012)^-$ decay to $\Xi^-\pi^0\bar K^0$ and $\Xi^0\pi^0K^-$ are equal, and the latter two are half the former two. 
Accounting for detection efficiencies and the branching fractions for how we observe the decays of the final-state particles,
the proportions of events we should see in each final state are 86.4\% $\Xi^-\pi^+K^-$, 7.4\% $\Xi^0\pi^-\bar K^0$, 3.9\% $\Xi^0\pi^0K^-$, and 2.3\% $\Xi^-\pi^0\bar K^0$. In the fit, we assume the yields for the final states have these proportions.
The yields for $\Omega(2012)^-$ two-body decay are floated in the fit.

All signal shapes are described using Flatt$\acute {\rm e}$-like functions~\cite{094028}, convolved with Gaussian functions to model detector resolutions---1.5, 2.5, 2.3, 2.8, 1.7, and 2.1 MeV for $\Omega(2012)^-\to\Xi^-\pi^+K^-$, $\Xi^-\pi^0K^0_S$, $\Xi^0\pi^-K^0_S$, $\Xi^0\pi^0K^-$, $\Xi^-K^0_S$, and $\Xi^0K^-$, determined from inspection of simulated data, and multiplied by efficiency functions linearly dependent on mass.
The Flatt$\acute {\rm e}$-like function is

\vspace{-0.5cm}
\begin{equation}\label{eq:1}
\small
T_n(M) \equiv \frac{g_nk_n(M)}{|M-m_{\Omega(2012)^-}+\frac{1}{2}\sum\limits_{j=2,3}g_j[\kappa_j(M)+ik_j(M)]|^2},
\end{equation}
where $g_n$ is the effective coupling of $\Omega(2012)^-$ to the $n$-body final state, $\kappa_n$ and $k_n$ parameterize the real and imaginary parts of the $\Omega(2012)^-$ self-mass,
and $M$ is $M(\Xi\bar K)$ or $M(\Xi\pi\bar K)$.
The forms of $\kappa_n$ and $k_n$ are found in Eqs.~(47) and~(48) in Ref.~\cite{094028} with mass-dependent widths.
Such modeling of the $\Omega(2012)^-$ shape, not included in the previous analysis~\cite{032006}, improves the signal-background separation and hence increases the signal yield. For $\Omega(2012)^-\to \Xi\pi\bar K$, we assume the decay proceeds only through the $\Xi(1530)$ resonance.
When we include a nonresonant component in the fit, its yield is consistent with zero.

Background events in the $\Xi\pi\bar K$ mass distributions are modeled by the same threshold function from Ref.~\cite{032006}, but with a threshold of 1.95 GeV. Background events in the $\Xi\bar K$ mass distributions are modeled by second-order polynomials.

The fit results are shown as blue lines in Fig.~\ref{fig2} and the yields are listed in Table~\ref{tab2}.
The statistical significance for $\Omega(2012)^-\to \Xi(1530)\bar K\to \Xi\pi \bar K$ is 5.4~$\sigma$, calculated from the logarithm of a ratio of likelihoods~\cite{60}, $-2{\rm ln}({{\cal L}_0}/{{\cal L}})$, which is 33,
accounting for the difference in the number of degrees of freedom (2),
where ${\cal L}_0$ and ${\cal L}$ are the maximized likelihoods without and with the signal components~\cite{60}.
Including systematic uncertainties, the significance is $5.2~\sigma$. The significance of the two-body decays is 8.8~$\sigma$.
The correlations of the yields are 0.136 for $\Omega(2012)^-\to \Xi^-\bar K^0$ and $\Omega(2012)^-\to \Xi^0K^-$, 0.142 for $\Omega(2012)^-\to \Xi^-\bar K^0$ and $\Omega(2012)^-\to\Xi\pi\bar K$, and 0.167 for $\Omega(2012)^-\to \Xi^0K^-$ and $\Omega(2012)^-\to\Xi\pi\bar K$.
The signal yields for $\Omega(2012)^-\to \Xi^0K^-$ and $\Omega(2012)^-\to \Xi^-K^0_S$ are consistent with those from Ref.~\cite{052003}. The mass of $\Omega(2012)^-$ is

\vspace{-0.2cm}
\begin{equation}\label{eq:add1}
m_{\Omega(2012)^-}=(2012.5\pm0.7\pm0.5)~{\rm MeV}.
\end{equation}The values of $g_3$ and $g_2$ are
\begin{equation}\label{eq:add2}
\begin{aligned}
&g_3=(38.9^{+31.1}_{-38.9}\pm9.0)\times10^{-2},\\
&g_2=(1.7^{+0.3}_{-0.3}\pm0.3)\times10^{-2}.
\end{aligned}
\end{equation}The value of $g_3/g_2$ is
\begin{equation}\label{eq:add3}
\begin{aligned}
g_3/g_2=(22.9^{+17.9}_{-22.4}\pm2.2).
\end{aligned}
\end{equation}Where two uncertainties are given, the first is statistical and the second systematic.
The effective couplings essentially determine the width of the $\Omega(2012)^-$.
The shape of the signal does not change greatly with $g_3$. Even if $g_3$ is very close to zero, 
the Flatt$\acute {\rm e}$-like function can still describe the $M(\Xi \pi \bar K)$ distribution in the data. 
We do not have enough data to precisely determine $g_3$ in the limited range where it has an effect, so its relative statistical uncertainty is much larger than those of $m_{\Omega(2012)^-}$ and $g_2$.

\begin{figure}[t]
\centering
\includegraphics[width=6cm]{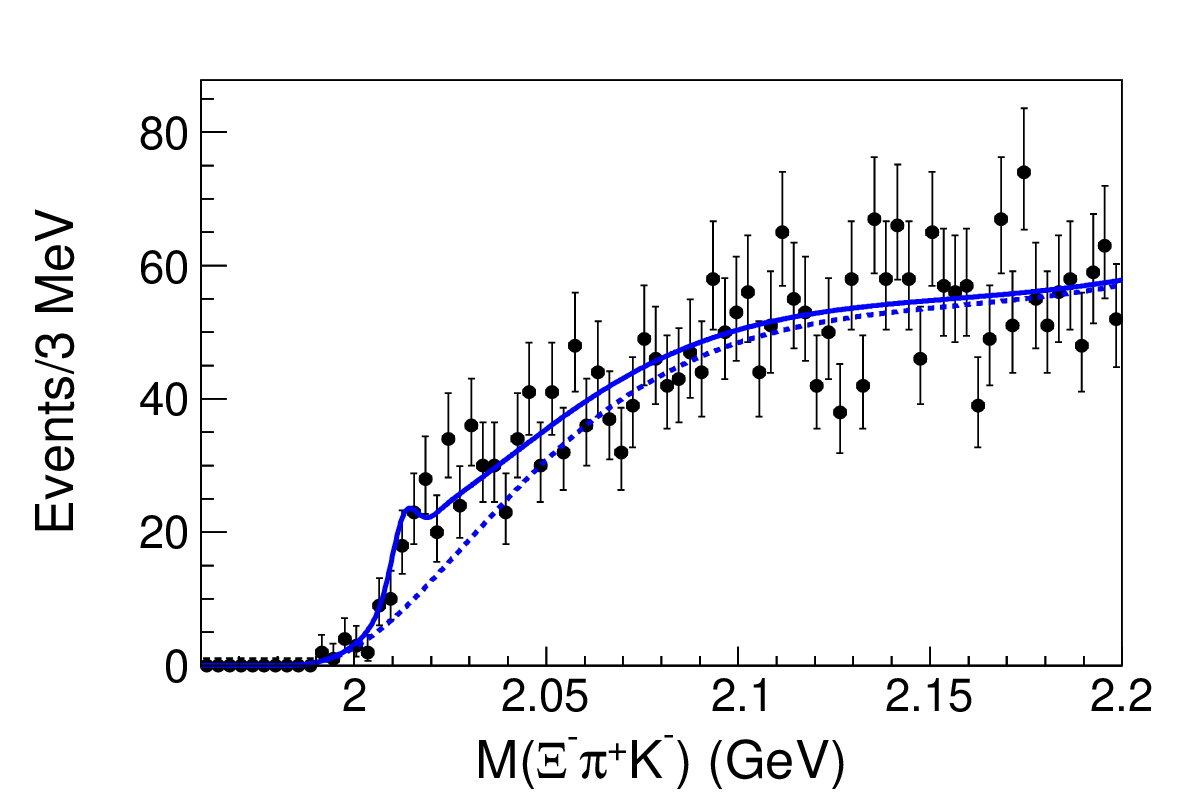}
\caption{The $\Xi^-\pi^+K^-$ mass distribution for events with $M(\Xi\pi)$ $\in$ $[1490,1530]$\ MeV in data. The solid curve is the best fit and the dashed curve is the background.}\label{cs2}
\end{figure}

The previous analysis found only $22\pm13$ signal events with a $\chi^2$ per degree of freedom of 37.3/12~\cite{032006}. 
The signal yield increases greatly with the new model.
This supports that we now more accurately model $\Omega(2012)^- \to \Xi(1530) \bar K \to \Xi \pi \bar K$ and that the selection criteria and fit functions based on this model are reasonable.

For comparison, we also fit the $M(\Xi^-\pi^+K^-)$ distribution using a Breit-Wigner function convolved with a Gaussian resolution function for the $\Omega(2012)^-$ signal shape, the same used in previous analysis~\cite{032006}. This fit function does not describe the data very well especially, in the $\Omega(2012)^-$ signal region. 
The $\chi^2$ per degree of freedom of the result of the fit is 16.2/12. In the fit using our new model, it is much smaller, 4.8/11.

We also fit our new model to data selected with the previous mass requirement, $M(\Xi\pi)$ $\in$ $[1490,1530]$ MeV, shown in Fig.~\ref{cs2}, to demonstrate the improvement of the new mass requirement. We fix $m_{\Omega(2012)^-}$, $g_2$, and $g_3$ to the central values in equations~\ref{eq:add1} and~\ref{eq:add2}.
The yield for $\Omega(2012)^-\to\Xi(1530)^0K^-\to\Xi^-\pi^+K^-$ is $212\pm53$ with a $\chi^2$ per degree of freedom of 23.0/14.
The yield reduction is consistent with the 23\% loss expected from inspection of simulated data.

\begin{table}[t]
\centering
\scriptsize
\caption{The efficiencies ($\varepsilon$), signal yields ($Y$), and branching-fraction products ($b$) for all channels.}\label{tab2}
\begin{tabular}{lrrr} \hline
Mode & $\varepsilon$ (\%) & $Y$ & $b$ (\%)\\\hline
$\Omega(2012)^-\to \Xi(1530)^0K^-\to\Xi^-\pi^+K^-$ & $6.97\pm0.07$ & $267\pm60$ & 64 \\
$\Omega(2012)^-\to \Xi(1530)^-\bar K^0\to\Xi^-\pi^0\bar K^0$ & $1.06\pm0.01$ & $7\pm2$& 22\\
$\Omega(2012)^-\to \Xi(1530)^-\bar K^0\to\Xi^0\pi^-\bar K^0$ & $1.74\pm0.02$ & $23\pm5$ & 22\\
$\Omega(2012)^-\to \Xi(1530)^0K^-\to\Xi^0\pi^0K^-$ & $0.63\pm0.01$ & $12\pm3$ & 62 \\
$\Omega(2012)^-\to \Xi^0K^-$ & $4.00\pm0.04$ & $242\pm40$ &63\\
$\Omega(2012)^-\to \Xi^-\bar K^0$ & $15.5\pm0.16$ & $293\pm65$ &22 \\
\hline
\end{tabular}
\end{table}

\section{Branching-fraction ratio}\label{Sec1}

The numerator in ${\cal R}^{\Xi \pi \bar K}_{\Xi \bar K}$ is the sum of the branching fractions for the four decays $\Omega(2012)^- \to \Xi(1530) \bar K$ with subsequent $\Xi(1530)$ decay resulting in the final states $\Xi^-\pi^+K^-$, $\Xi^{-}\pi^{0}\bar K^{0}$, $\Xi^{0}\pi^{-}\bar K^{0}$, and $\Xi^0\pi^0K^-$.
Assuming isospin symmetry,

\vspace{-0.3cm}
\begin{equation}
\small
\begin{aligned}
{\cal R}^{\Xi \pi \bar K}_{\Xi \bar K} = \frac{3\times\BR(\Omega(2012)^-\to \Xi(1530)^0K^-\to\Xi^- \pi^+K^-)}{\BR(\Omega(2012)^-\to \Xi^{-}\bar K^{0})+\BR(\Omega(2012)^-\to \Xi^{0}K^-)},
\end{aligned}
\end{equation}
where

\vspace{-0.5cm}
\begin{equation}
\mathcal{B}(X) = \frac{Y(X)}{\varepsilon(X) \, b(X) \, N_{\Omega(2012)^-}},
\end{equation}
with $Y(X)$ the yield of channel $X$, $\varepsilon(X)$ the efficiency to detect it, $b(X)$ the product of branching fractions for the final-state-particle detection modes needed to detect channel $X$ (a subset of $\Xi^{+} \to \Lambda^0 \pi^{+}$, $\Xi^{0} \to \Lambda^0 \pi^{0}$, $\Lambda^0 \to p \pi^-$, $\bar K^0 \to \pi^+\pi^-$, and $\pi^0 \to \gamma \gamma$), and $N_{\Omega(2012)^-}$ the number of $\Omega(2012)^-$ produced at Belle, which cancels in the ratio. The yields, efficiencies, and branching-fraction products (calculated with values from \cite{PDG}) are listed in Table~\ref{tab2}. We obtain

\vspace{-0.5cm}
\begin{alignat}{1}
{\cal R}^{\Xi\pi{\bar K}}_{\Xi {\bar K}} = 0.99\pm0.26\pm0.06.
\end{alignat}
Since the $\Omega(2012)^-$ and the intermediate state, $\Xi(1530)\bar{K}$, have similar masses, the branching-fraction ratio depends more strongly on $m_{\Omega(2012)^-}$ and $g_2$ than on $g_3$. We can therefore measure it more precisely than $g_3/g_2$.

\section{Systematic uncertainties}

Systematic uncertainties on the branching-fraction ratio are summarized in Table~\ref{systematic2}. Systematic uncertainty on the detection efficiency comes from tracking (0.35\% per track), particle identification (1.3\% per kaon and 1.1\% per pion), $K^0_S$ selection (2.2\%)~\cite{171801}, and $\pi^0$ reconstruction (4\%)~\cite{072004}.
Uncertainties from $\Lambda^0$ detection cancel.
The particle-identification uncertainties are added linearly.
By including these uncertainties in the fit, we obtain the total systematic uncertainty from detection efficiency, 3.1\%.

The masses and widths of $\Lambda^0$, $\Xi^-$, $\Xi^0$, and $\Xi(1530)^0$ and the mass resolution of $\Omega(2012)^-$ are fixed in our fit.
We change the masses and widths each by one unit of their uncertainty, refit, and take the differences in yields as systematic uncertainties.
Simulation usually understimates the resolutions of mass peaks within 10\%~\cite{052003}.
We enlarge the resolution by 10\% and take the resulting change as a conservative systematic uncertainty.
These uncertainties are added in quadrature to obtain a systematic uncertainty due to resonance parameters, 5.3\%.

Fit-model uncertainties are estimated using simulated pseudoexperiments, with events distributed according to the mass spectra observed in data.
We repeat this numerous times and find that the mean of the results of these fits is consistent with the fit to data. This cross checks the fit's stability.

When we allow the proportions of the three-body decays to break isospin symmetry by up to 10\%, the fitted
yields change by less than 1\%. 
So uncertainty on the branching-fraction ratio from isospin-symmetry is negligible.
The statistical uncertainties on the efficiencies determined from simulation and the uncertainties from the branching fractions for $\Xi\to\Lambda^0\pi$, $K^{0}_{S}\to \pi^{+}\pi^{-}$, and $\pi^0\to\gamma\gamma$ are also negligible~\cite{PDG}.

Systematic uncertainties on the $\Omega(2012)^-$ mass and effective couplings are summarized in Table~\ref{systematic2}.
The uncertainties from the input resonance parameters are 0.1 MeV for the $m_{\Omega(2012)^-}$, 0.034 for $g_3$, and 0.001 for $g_2$.
We change the fit ranges and change the background shapes to higher-order polynomials and take the differences of 0.5 MeV on the $m_{\Omega(2012)^-}$, 0.083 on $g_3$, and 0.002 on $g_2$ as systematic uncertainties. 
We allow the proportions of the three-body decays to break isospin symmetry by up to 10\% and take the differences of 0.1 MeV on the $m_{\Omega(2012)^-}$, 0.002 on $g_3$, and 0.001 on $g_2$ as systematic uncertainties.
The statistical uncertainties on the efficiencies determined from simulation and the uncertainties from the branching fractions are negligible in determinations of the $\Omega(2012)^-$ mass and effective couplings.
The total systematic uncertainties on the $m_{\Omega(2012)^-}$, $g_3$, and $g_2$ are 0.5 MeV, 0.090, and 0.003, respectively.

\begin{table}[htbp]
\centering
\scriptsize
\caption{Systematic uncertainties on the branching-fraction ratio, $\Omega(2012)^-$ mass, and effective couplings.}\label{systematic2}
\begin{tabular}{lrrrr} \hline
Source & ${\cal R}^{\Xi\pi K}_{\Xi \bar K}$ & $m_{\Omega(2012)^-}$ [MeV] & $g_3$ & $g_2$ \\\hline
Detection efficiency & 3.1 & --- & --- & --- \\
Resonance parameters & 5.3 & 0.1 & 0.034 & 0.001 \\
Fit model & --- & 0.5 & 0.083 & 0.002 \\
Isospin symmetry & --- & 0.1 & 0.002 & 0.001 \\\hline
Total & 6.1 & 0.5 & 0.090 & 0.003 \\\hline
\end{tabular}
\end{table}

\section{Summary}

We report the first observation of $\Omega(2012)^-\to\Xi(1530)\bar K\to \Xi\pi\bar K$,
using $\Upsilon(1S)$, $\Upsilon(2S)$, and $\Upsilon(3S)$ data from Belle. The mass of the $\Omega(2012)^-$ is $(2012.5\pm0.7\pm0.5)~{\rm MeV}$; its effective coupling to $\Xi(1530)\bar{K}$ is $(39^{+31}_{-39}\pm9)\times10^{-2}$; and its effective coupling to $\Xi\bar{K}$ is $(1.7\pm0.3\pm0.3)\times10^{-2}$.
Assuming isospin symmetry, the ${\cal R}^{\Xi\pi{\bar K}}_{\Xi {\bar K}}$ is $0.99\pm0.26\pm0.06$.

\section{ACKNOWLEDGMENTS}

We warmly thank Prof.\ Bing-Song Zou and Prof.\ Ju-Jun Xie for valuable and helpful discussions.
This work, based on data collected using the Belle detector, which was
operated until June 2010, was supported by 
the Ministry of Education, Culture, Sports, Science, and
Technology (MEXT) of Japan, the Japan Society for the 
Promotion of Science (JSPS), and the Tau-Lepton Physics 
Research Center of Nagoya University; 
the Australian Research Council including grants
DP210101900, 
DP210102831, 
DE220100462, 
LE210100098, 
LE230100085; 
Austrian Federal Ministry of Education, Science and Research and
Austrian Science Fund (FWF) No.~P~31361-N36;
National Key R\&D Program of China under Contracts No.~2022YFA1601903 and No.~2024YFA1610503,
National Natural Science Foundation of China and research grants
No.~12475076,
No.~11575017,
No.~11761141009, 
No.~11705209, 
No.~11975076, 
No.~12135005, 
No.~12150004, 
No.~12161141008, 
and
No.~12175041, 
and Shandong Provincial Natural Science Foundation Project ZR2022JQ02;
the Czech Science Foundation Grant No. 22-18469S;
Horizon 2020 ERC Advanced Grant No.~884719 and ERC Starting Grant No.~947006 ``InterLeptons'' (European Union);
the Carl Zeiss Foundation, the Deutsche Forschungsgemeinschaft, the
Excellence Cluster Universe, and the VolkswagenStiftung;
the Department of Atomic Energy (Project Identification No. RTI 4002), the Department of Science and Technology of India,
and the UPES (India) SEED finding programs Nos. UPES/R$\&$D-SEED-INFRA/17052023/01 and UPES/R$\&$D-SOE/20062022/06; 
the Istituto Nazionale di Fisica Nucleare of Italy; 
National Research Foundation (NRF) of Korea Grant
Nos.~2016R1-D1A1B-02012900, 
2018R1-A6A1A-06024970, 
2021R1-A6A1A-03043957, 
2021R1-F1A-1060423, 
2021R1-F1A-1064008, 
2022R1-A2C-1003993, 
2022R1-A2C-1092335, 
RS-2022-00197659, 
RS-2023-00208693; 
Radiation Science Research Institute, Foreign Large-size Research Facility Application Supporting project, the Global Science Experimental Data Hub Center, the Korea Institute of Science and Technology Information (K24L2M1C4) and KREONET/GLORIAD; 
the Polish Ministry of Science and Higher Education and 
the National Science Center;
the Ministry of Science and Higher Education of the Russian Federation
and the HSE University Basic Research Program, Moscow; 
University of Tabuk research grants
S-1440-0321, S-0256-1438, and S-0280-1439 (Saudi Arabia);
the Slovenian Research Agency Grant Nos. J1-9124 and P1-0135;
Ikerbasque, Basque Foundation for Science, and the State Agency for Research
of the Spanish Ministry of Science and Innovation through Grant No. PID2022-136510NB-C33 (Spain);
the Swiss National Science Foundation; 
the Ministry of Education and the National Science and Technology Council of Taiwan;
and the United States Department of Energy and the National Science Foundation.
These acknowledgements are not to be interpreted as an endorsement of any
statement made by any of our institutes, funding agencies, governments, or
their representatives.
We thank the KEKB group for the excellent operation of the
accelerator; the KEK cryogenics group for the efficient
operation of the solenoid; and the KEK computer group and the Pacific Northwest National
Laboratory (PNNL) Environmental Molecular Sciences Laboratory (EMSL)
computing group for strong computing support; and the National
Institute of Informatics, and Science Information NETwork 6 (SINET6) for
valuable network support.


\begin{thebibliography}{**}
\bibitem{unit} We use units in which the speed of light is 1 and unitless.
\end{thebibliography}
\end{document}